\documentclass[conference]{IEEEtran}
\IEEEoverridecommandlockouts
\usepackage{cite}
\usepackage{amsmath,amssymb,amsfonts}
\usepackage{algorithmic}
\usepackage{graphicx}
\usepackage{textcomp}
\usepackage{xcolor}
\usepackage{url}
\def\BibTeX{{\rm B\kern-.05em{\sc i\kern-.025em b}\kern-.08em
    T\kern-.1667em\lower.7ex\hbox{E}\kern-.125emX}}
\begin{document}

\title{TLS Beyond the Broker: Enforcing Fine-grained Security and Trust in Publish/Subscribe Environments for IoT\\
{\footnotesize \textsuperscript{*}Note: This is a Pre-print version. The paper is part of the 17th International Workshop on Security and Trust Management (STM2021).}
}

\author{\IEEEauthorblockN{Korbinian Spielvogel}
\IEEEauthorblockA{\textit{Passau Institute of Digital Security} \\
\textit{University of Passau}\\
D-94032 Passau \\
office@sec.uni-passau.de}
\and
\IEEEauthorblockN{Henrich C. Pöhls}
\IEEEauthorblockA{\textit{Passau Institute of Digital Security} \\
\textit{University of Passau}\\
D-94032 Passau \\
office@sec.uni-passau.de}
\and
\IEEEauthorblockN{Joachim Posegga}
\IEEEauthorblockA{\textit{Passau Institute of Digital Security} \\
\textit{University of Passau}\\
D-94032 Passau \\
office@sec.uni-passau.de}
}

\maketitle

\begin{abstract}
Message queuing brokers are a fundamental building block of the Internet of Things, commonly used to store and forward messages from publishing clients to subscribing clients. Often a single trusted broker offers secured (e.g. TLS) and unsecured connections but relays messages regardless of their inbound and outbound protection. Such mixed mode is facilitated for the sake of efficiency since TLS is quite a burden for MQTT implementations on class-0 IoT devices.
Such a broker thus transparently interconnects securely and insecurely connected devices; we argue that such mixed mode operation can actually be a significant security problem: Clients can only control the security level of their own connection to the broker, but they cannot enforce any protection towards other clients.
We describe an enhancement of such a publish/subscribe mechanism to allow for enforcing specified security levels of publishers or subscribers by only forwarding messages via connections which satisfy the desired security levels. For example, a client publishing a message over a secured channel can instruct the broker to forward the message exclusively to subscribers that are securely connected. We prototypically implemented our solution for the MQTT protocol and provide detailed overhead measurements.

\end{abstract}
\section{Introduction}

Internet of Things (IoT) environments often rely on the exchange of messages between plentiful devices; the multitude of possible communication flows has led to the introduction of message passing systems, better known as publish/subscribe systems or message \textit{brokers}. 
Quite obviously, a secure connection between message senders (\textit{publishers}) and message receivers (\textit{subscribers}) is important for many application areas. 
The security of such systems clearly also relies on the security of the message exchange facilitated by the broker~\cite{hivemq_mqtt_security_fundamentals:2015}. 
Since  IoT devices are quite heterogeneous and have very different  capabilities, e.g. some are not always online or have limited computational or battery power~\cite{kaspersky:2018}, devices might also have very different needs in terms of their security goals / requirements in terms of confidentiality or origin authentication of messages.
A proper IoT system design should therefore allow devices to specify security needs and  enforce them towards other devices.

The role of the message broker is to facilitate data exchange between loosely coupled endpoints by providing  message queuing, which decouples sending devices from receiving ones. 
Thus, there is usually no direct relationship between the publisher and the subscriber and one must rely on the broker to enforce a a given security level. We implemented such a broker as a prototype, based on the  Message Queuing Telemetry Transport (MQTT) protocol, one of the most common protocols in the IoT \cite{mqtt_popularity:2019}.
MQTT enables efficient communication even with limited computational power and constrained resources \cite{mqtt5:oasis}.
Essentially, the broker allows clients to publish (send) messages belonging to a specific category, usually referred to as \textit{topic}. 
A published message is then stored and forwarded by the broker to all clients that signalled their interest by subscribing to that message's topic.
The participating entities are usually refered to as \textit{publishers} and \textit{subscribers}.
If messages contain privacy-sensitive data, then exchanging these messages without suitable protection can easily be a significant security problem~\cite{bookEngineeringSecureIoTSystems}. Furthermore, many devices autonomously send information to the manufacturer's servers which allows for insights into the physical activities at the location where the devices are in use \cite{felixkorbihenrich2020}. 

\begin{figure}[ht!]
    \centering
    \includegraphics[width=0.5\textwidth]{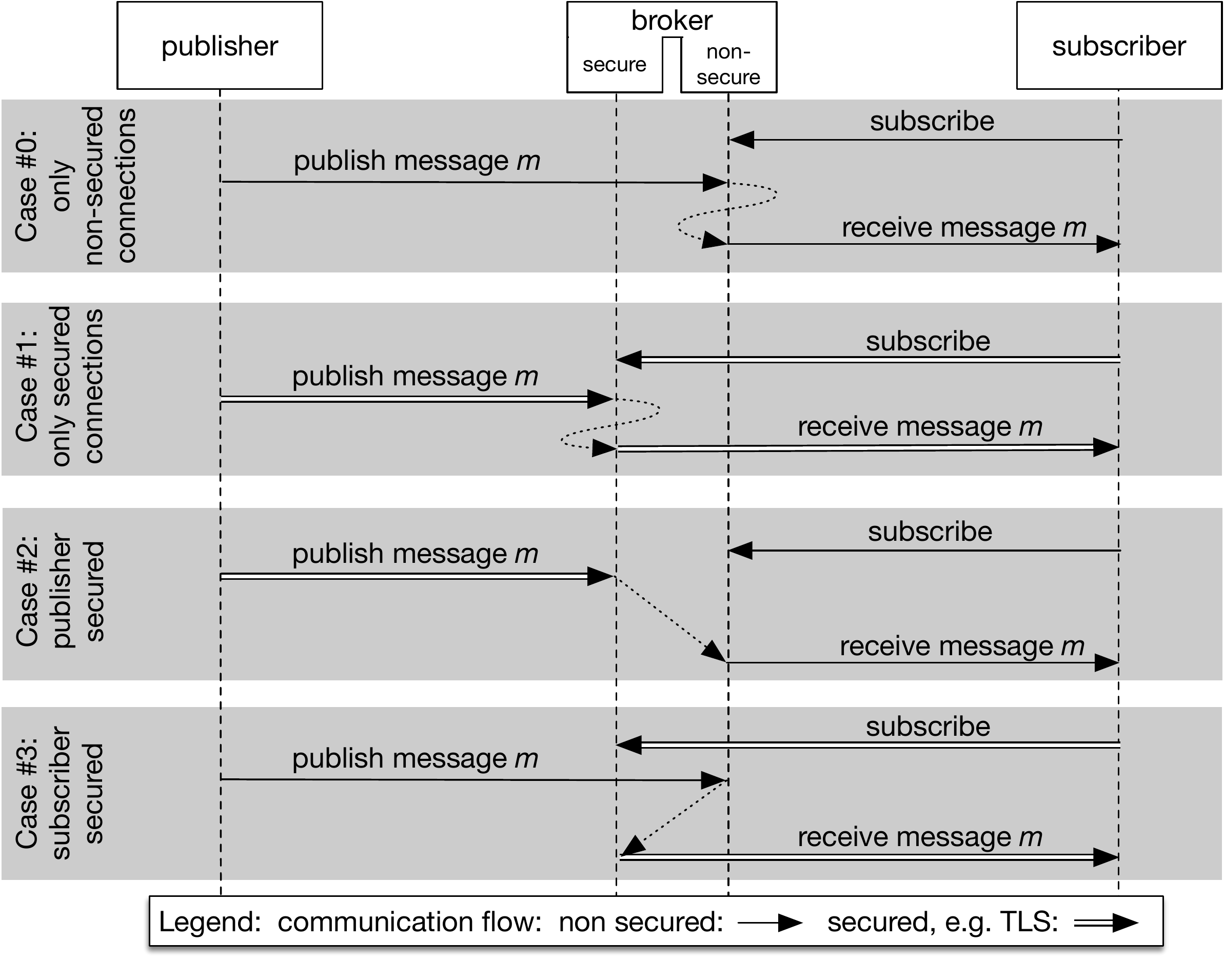}
    \caption{The four different cases of message forwarding when relaying messages between clients connecting secured and non-secured to the broker}
    \label{fig:my_label}
\end{figure}

Two attacks are possible with message queuing systems that do not secure the communication between their senders and receivers (see Fig. \ref{fig:my_label}), which is the case with a plain usage of the MQTT protocol: Firstly, confidentiality can be compromised and unauthenticated third parties can  read arbitrary messages.  Second, unauthenticated third parties might even create their own messages and inject them into the system, thus compromising integrity. 
The MQTT protocol is known to have several critical security issues in its default configuration, i.e.:
if implemented/configured without additional security mechanisms~\cite{security_issues_mqtt}. 
These problems have mainly arisen over time, as the protocol at the time of development had other objectives than security, simply because it was not considered to be significant at that time \cite{hivemq_introduction:2015} as one relied on physical security or other protection mechanisms. This has changed significantly over time. Application areas like automotive, smart home, logistics or transportation sectors, where MQTT is widely used, now consider security as one of the primary objectives. Compromising integrity and confidentiality of messages by third parties in the message queuing system can therefore have fatal consequences. \cite{analysis_of_mqtt_vulnerabilities:2018}

The standard solution here is to enable encryption and authentication between the clients (subscribers and publishers) and the broker based on the well established~\footnote{see \url{https://transparencyreport.google.com/} or \url{https://telemetry.mozilla.org/}} and scrutinised~\cite{tls_scrutinise:2016,tls13-proof} 
 transport layer security (TLS) protocol. The clients can then authenticate the broker and the broker can authenticate the clients using certificates or usernames and passwords. 
The client's secure connection will then end at the broker, which is generally seen as trusted. However, some IoT clients are too restricted to use TLS-secured connections, either not for all interactions, or not at all; many message queuing systems  allow clients to connect  over secured (e.g. using TLS) as well as non-secured connections to a broker. Such a broker might then  relay messages between clients connected with a different level of security and forward a message received over TLS via an non-encrypted connection  (see Fig.~\ref{fig:my_label}). The securely connected client of a broker cannot enforce secure connections on other clients and prevent forwarding over insecure connections as shown in case 2 and 3 of Fig.~\ref{fig:my_label}.
One could isolate non-secured connected devices from securely-connected ones, but this would  either disable sensible message flows, or require many more devices to use TLS, which is costly in terms of compute power or run-time.

The rest of this paper is organised as follows: We propose  a solution that enables hybrid security levels among  connected clients, so clients can connect over secured as well as non-secured connections; we achieve this by extending the trusted broker to enforce a certain level beyond the initial connection on a fine-grained, per message or per topic, policy. A thorough analysis of the  problem  underlying the MQTT protocol is presented in Section~\ref{sec:problem}; we compare our novel, broker-enforced solution to related work in Section~\ref{sec:related}. The solution was prototyped for MQTT, and we argue in Section \ref{sec:ouridea} that the concept of an  expansion to enforce security levels beyond the broker does generally not require to change the existing clients. 
Adjustments to the MQTT clients would of course allow optimisations in signalling the intended security level.  We  measured the overhead of TLS, which is saved when the client could decide to not facilitate a TLS connection for certain messages or topics: Section~\ref{sec:speed} presents  measurements of the overhead in runtime of TLS for a device from the smallest class\footnote{Class-0 is defined as the group of IoT devices with less than 10KB RAM (Data size) and less than 100KB ROM (Code size) ~\cite{iotdevice_classes:2016}} of IoT devices ~\cite{iotdevice_classes:2016}, i.e. a Wemos D1 mini with the ESP8266EX chip running at 80/160MHz and only 50kB of RAM and up to 16MB external SPI flash ~\cite{espressif_documentation}. 
In Section~\ref{sec:discussion} we discuss how our approach relates to prior work, like end-to-end or multi-lateral security and conclude in Section~\ref{sec:conclusion}.

\section{Problem: No differentiation of incoming or outgoing security goals}
\label{sec:problem}

We will start with recalling some details of the MQTT protocol and its entities  before we describe the general problem of the missing differentiation between the security goals of clients in hybrid broker environments in Section~\ref{sec:hybridproblem}. 


\subsection{MQTT protocol and its entities}
\label{sec:mqttbackground}
An MQTT client is in principle any device that speaks MQTT over a TCP/IP stack, i.e. that runs an MQTT library and can connect to an MQTT broker via a network. Furthermore, MQTT clients are distinguished between MQTT publisher and MQTT subscriber. Whether a client is a publisher, a subscriber or even both depends on whether the client sends messages (publisher) or receives messages (subscriber). The messages here are assigned to a certain topic, which categorizes the messages according to their subject. In MQTT, the term "topic" refers to a UTF-8 string that can consist of several different levels which are separated by a forward slash ("/"). \cite{hivemq_connect_establishment:2019}\newline
For example, a smart thermometer located in the kitchen of a house could act as an MQTT publisher and send the current temperature of the kitchen as an MQTT message. The corresponding topic for this would be, for example, "home/kitchen/temperature". In turn, another device, such as a home owner's mobile phone, could act as an MQTT subscriber, monitoring the important values of the house, such as temperature, humidity, electricity etc. To do so, the MQTT subscriber would have to subscribe to the topic "home" to receive all messages that have "home" as the first topic-level. \newline
At this point, it should be noted that in the remaining parts of the paper we will use the term "TCP client" as a synonym for any MQTT client that uses plain TCP as the underlying transport protocol and does not implement any additional security measurements to address the vulnerabilities of MQTT. Any client that uses TLS in addition to TCP will be referred to as a "TLS client". 

The MQTT Broker is certainly the most important component of the MQTT protocol. It constitutes the interface between the MQTT clients and manages the exchange of messages between them. The MQTT broker thereby ensures that the messages of a certain topic sent by an MQTT publisher are forwarded to the MQTT subscribers which have subscribed to this topic. In addition, the broker also handles the management of the persistent sessions as well as the authentication and authorization of the clients. Since this component is often directly exposed to the Internet, it is particularly important to ensure security, scalability, monitoring and failure resistance of the broker. \cite{hivemq_connect_establishment:2019}

The minimum requirement for an MQTT system to be functional and useful at the same time is to include (i) a central broker, which forwards the incoming messages of the publishers to the respective subscribers (ii) a subscriber, that receives messages of at least one topic, and (iii) a publisher, that sends a message to that topic. The subscriber and the publisher may also be the same client.

For simplicity the message flow between one publisher, one broker and one subscriber in Fig.~\ref{fig:my_label} shows only the message and implicitly assumes that publisher and subscriber use the same topic. 
Moreover, Fig.~\ref{fig:my_label} is simplified by having only one publisher and only one subscriber; in general the publish/subscribe architecture is best fitting if there are one-to-many or even many-to-many relations, e.g. \textit{m} subscribers are listening to the messages of \textit{n} publishers. 

In reality the MQTT brokers and the protocol are built to scale up to large amounts of clients, subscribers as well as publishers~\cite{mqtt_survey:2020,mqtt5:oasis}.
Last but not least, Fig.~\ref{fig:my_label} is simplified in respect to the MQTT protocol itself, as MQTT would actually require an initial handshake (\texttt{CONNECT} message) and has acknowledgment messages that indicate the successful reception of commands for connection, subscription and publication (\texttt{CONACK}, \texttt{SUBACK} and \texttt{PUBACK}).

\subsection{Trust assumptions among subscribers, publishers and brokers }
In a publish/subscribe message queuing system the clients trust the broker to process the messages correctly, i.e. the publisher trusts the broker to forward the message to all interested subscribers.
Subscribers trust the broker to deliver them all messages of their indicated interest, e.g. the topic they subscribed to.  

To increase the security each client can secure the connection to the broker. 
As discussed this can be done using the MQTT protocol over TLS.
Current implementations of MQTT brokers allow to enable TLS and once configured the broker enforces TLS for all clients and strictly rejects all other connection types, such as plain TCP. 
Devices with even fewer resources than the client in our experiments (e.g., Class-0 IoT devices as defined by \cite{iotdevice_classes:2016}) could thus become excluded from the system, as they are not able to allocate sufficient resources for performing a full TLS handshake and are therefore not TLS-capable. 
For MQTT in particular, it is also conceivable that many of the devices do not require security-sensitive message exchanges and thus have no need for a secure communication channel with the broker. 
Hence, it would be an unnecessary additional consumption of already rather scarce resources if these devices all had to use TLS. 
This is why the broker could be deployed in a hybrid or mixed mode, supporting both secured and non-secured connections of its clients.

\subsection{Hybrid or mixed mode brokers and TLS on small devices}
\label{sec:speed}
To not limit the connection to TLS-capable devices only --as discussed above-- many MQTT brokers, such as HiveMQ or Mosquitto, allow the configuration of multiple so-called listeners. Each listener enables the specification of one out of several different connection types \cite{hivemq_listeners:2020}. 
For instance, it is possible to configure a broker such that it accepts connections from clients using both plain TCP (port 1883) and TLS (port 8883) simultaneously.
A client can then decide which connection type is most suitable for its purposes. 
Providing the possibility for clients to choose the connection type independently is particularly useful for clients with unreliable network connections, limited available bandwidth or highly limited resources. 
After all, in some cases it is more important to have a reliable, but possibly insecure connection in order to be able to transfer all data reliably.
~\cite{hivemq_listeners:2020}


The performance in terms of runtime for a TLS connection in comparison to a non-secured connection are shown in Table~\ref{tab:first_connection_eval}. 
On average, it takes roughly 335 times longer (more than 3 seconds) to establish a connection using TLS. The work of King and Awad ~\cite{iotdevice_classes:2016} even observed a delay of up to 24 seconds for certain IoT devices, even though their test setup used DTLS, which is a more efficient and resource-saving version of TLS. For many systems, such a performance loss is intolerable.
\begin{table}[tb]
	\centering
	\begin{tabular}{l|cc}
		\textbf{}                   & \multicolumn{1}{l}{\textbf{Insecure Connection}} & \multicolumn{1}{l}{\textbf{TLS Connection}} \\ \hline
		\textbf{Average Time}       & 9.54 ms                                          & 3207.53 ms                                  \\
		\textbf{Standard Deviation} & 3.69 ms                                          & 153.57 ms                                  \\
		\textbf{Minimum Time}       & 4.00 ms                                          & 3162.00 ms                                  \\
		\textbf{Maximum Time}       & 40.00 ms                                         & 5271.00 ms \\
		
	\end{tabular}
	\caption{Time to establish an initial connection between client and MQTT broker. Time measured from connection establishment (= sending \texttt{CONNECT}) till the \texttt{CONNACK} packet is received. TLSv1.2 with ECC (Curve25519) 256 bit keys and ciphersuite ECDHE-ECDSA-AES256-GCM-SHA384. MQTT broker is Eclipse Mosquitto (version 1.4.10) on Rasperrypi 3b+. Number of measurements: 500}
	\label{tab:first_connection_eval}
\end{table}
%
\subsection{Problem: No enforcement of incoming or outgoing clients' security goals beyond the broker}
\label{sec:hybridproblem}
We have identified the following problems of clients wanting to use mixed-mode brokers.
By mixed-mode brokers we refer to brokers offering non-secured as well as secured connections to their clients, but relaying messages based on the clients topics and not taking into account the security level of their client's connection.
For simplicity we differentiate only two security levels: non-secured and secured. 
In practice \textit{'secured'} means to send MQTT messages over a TLS connection while \textit{'non-secured'} uses just TCP. 
Thus, we either protect a message's confidentiality and integrity, or not.
It would be straightforward to extend our proposed solution to more than two and thus more fine grained security levels.

\begin{enumerate}
    \item[1$^{st}$] problem:\\ Client subscribing via a secured connection to a mixed-mode broker has no information about the security level of the connection of publishers, i.e. from a subscriber's perspective the received message is looking the same in case 1 (published secured) and case 3 (published non-secured) of Fig.~\ref{fig:my_label}.
    \item[2$^{nd}$] problem:\\  Client publishing via a secured connection to a mixed-mode broker can not enforce the security level of the connection of subscribers, i.e. from a publisher's perspective the sent message could either flow to the subscriber like in case 1 (secured) or as in case 2 (non-secured) of Fig.~\ref{fig:my_label}.
\end{enumerate}

Thus, in a mixed-mode environment there is a risk that the confidentiality or integrity of the message is only protected from the publisher to the broker, or only from the broker to the subscriber -- but it is never enforced beyond.
More flexibility would allow a client to enforce security levels beyond the broker if the application demands it, but also allow to not enforce this where needed:
\begin{enumerate}
    \item[1$^{st}$] scenario:\\  Publisher wants to extend security level enforcement beyond the broker to the subscribers
    \item[2$^{nd}$] scenario:\\ Subscriber wants to extend security level enforcement beyond the broker to the publishers 
    \item[3$^{rd}$] scenario:\\Publisher relaxes the security level for future communication beyond the broker to the subscribers
    \item[4$^{th}$] scenario:\\ Subscriber relaxes the security level for future communication beyond the broker to the publishers 
\end{enumerate}
This flexibility is currently not enforceable within the existing publish/subscribe model and not found in existing implementations.

\section{Related Work}
\label{sec:related}
The Message Queue Telemetry Transport protocol is a data transfer protocol developed to transfer data at the lowest possible cost. Additional mechanisms for protecting confidentiality and integrity were neglected accordingly~\cite{hivemq_tls_security_fundamentals:2015}. It replaces the typical end-to-end connection between devices in such a way that data producers and data users are decoupled. It is the origin of the Client-to-Broker architecture particularly used in the modern IoT. \cite{mqtt_history:2014}
The latest MQTT specification~\cite{mqtt5:oasis} defines the protocol as follows: 
\textit{"MQTT is a Client Server publish/subscribe messaging transport protocol. It is light weight, open, simple, and designed so as to be easy to implement. These characteristics make it ideal for use in many situations, including constrained environments such as for communication in Machine to Machine (M2M) and Internet of Things (IoT) contexts where a small code footprint is required and/or network bandwidth is at a premium."}. The protocol in general runs over TCP/IP or over any other network protocol that provides ordered, lossless and bi-directional connections~\cite{mqtt3_oasis}.

As shown in the previous chapters, the MQTT protocol includes a number of vulnerabilities in its default state. Many other authors have therefore analyzed the protocol for its vulnerabilities and presented methods to address these vulnerabilities. 
Andy et al. \cite{security_issues_mqtt} analyzed the number of freely available MQTT brokers with the help of Shodan. In addition, they demonstrated how attackers can violate the goals of data privacy, authentication, and data integrity in MQTT-based systems based on numerous scenarios. As a security measure against these numerous attack vectors, they propose the implementation of a security mechanism - specifically, the use of TLS. 
Harsha et al. \cite{analysis_of_mqtt_vulnerabilities:2018} also examined the possible attacks on MQTT based systems. In addition to the attacks already mentioned by Andy et al. \cite{security_issues_mqtt}, they illustrate how MQTT specific features such as wildcards can be exploited to subscribe to all topics simultaneously without knowing the specific topic name. Thus, they demonstrate how an attacker can disclose a maximum amount of information from an MQTT based system.  Furthermore, the authors prove that the integrated authentication mechanism of MQTT in the CONNECT message (= username and password authentication) is ineffective without additional protection mechanisms to ensure confidentiality, since the credentials are transmitted in plaintext and thus freely accessible to all attackers who are able to read or intercept network packets.

To address these vulnerabilities, Harsha et al. \cite{analysis_of_mqtt_vulnerabilities:2018} suggest the use of TLS or any mechanism for the encryption of the payload in combination with Access Control Lists (ACL) which define the access policies for a certain topic, i.e. which client can interact in which way with which topic. Finally, the authors demonstrated that TLS in combination with ACL can mitigate all of their defined attack-scenarios that previously violated the goals of confidentiality, authentication and integrity.
Closest to our approach is the TLS-proxy and firewall approach that was presented by Vučnik et al.~ \cite{mqtt_firewall:2019}. The authors implement a proxy that enforces TLS encrypted MQTT traffic for certain clients while allowing a non-secured plain TCP connection from others, e.g. those connected via the local network as local clients are seen as trustworthy. Their approach does not allow the clients to request the broker-mediated protection beyond the broker on a fine grained level of topics or messages, but just enforces it based on physical network location. It would be subject to future work to enhance our approach to take the physical network position into account when our broker is making the decision to forward messages.   

\section{Broker to enforce security levels beyond the broker}
\label{sec:ouridea}
\begin{figure}[t!]
    \centering
    \includegraphics[width=0.5\textwidth]{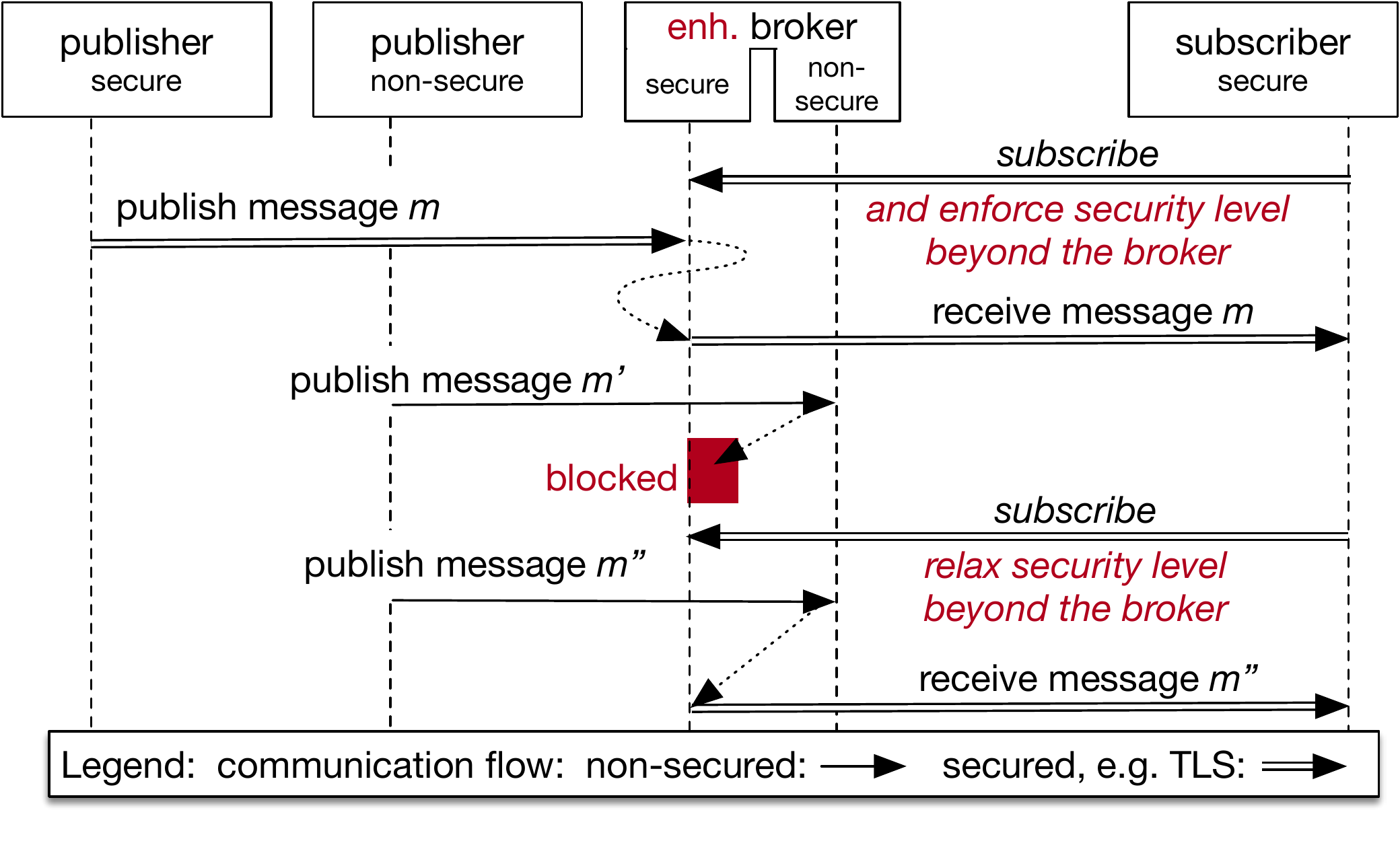}
    \caption{Our proposal: Trusted enhanced broker enforces publisher's security level}
    \label{fig:publisherenforcement}
\end{figure}

\begin{figure}[b!]
    \centering
    \includegraphics[width=0.5\textwidth]{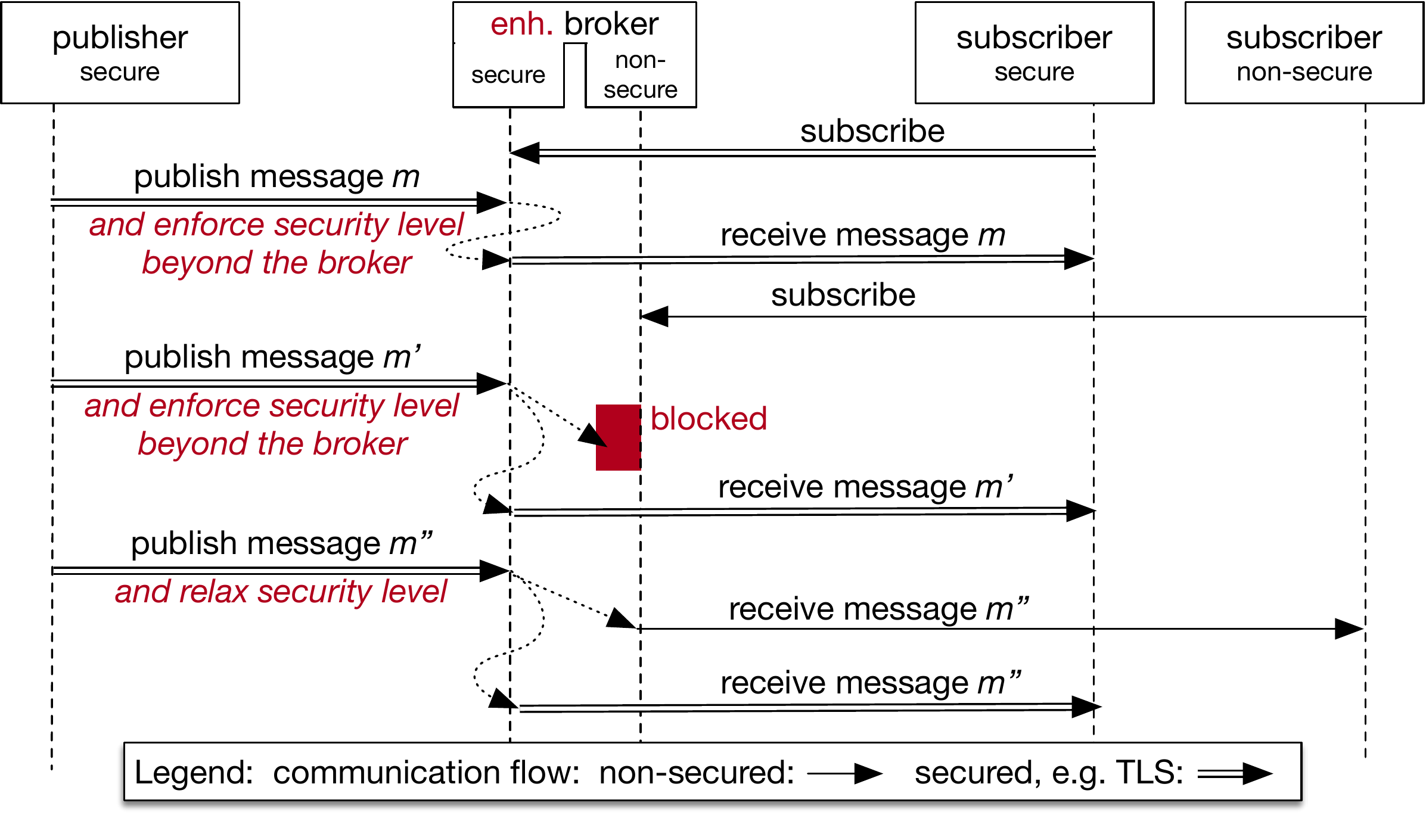}
    \caption{Our proposal: Trusted enhanced broker enforces subscriber's security level}
    \label{fig:subscriberenforcement}
\end{figure}

As we have presented the clients are in need to be able to specify the security levels that they would want to see enforced. 
In Fig.~\ref{fig:publisherenforcement} we have shown the intervention of the broker when the publishing client would like to enforce its security level of requiring secured connections beyond its own connection to the broker.
To enforce this the message \textit{m'} shown in Fig.~\ref{fig:publisherenforcement} is no longer forwarded by the broker to a subscriber that connected only over a non-secured channel.
The same would hold true for the enforcement of the security level of a subscriber, i.e. for a secure subscriber the broker would not forward --and thus the subscriber would not receive messages-- published by non-secured publishers, which is depicted for message \textit{m'} in Fig.~\ref{fig:subscriberenforcement}. 
We have termed the request to uphold the security level of the client to other clients beyond the broker as follows: \textit{broker-mediated }\textit{client-to-client} security.

%
%
\subsection{Client-to-client security mediated by the broker}
\label{sec:c2csec}
In the existing publish/subscribe model the brokers are trusted to forward messages.
In our solution the extended broker must be additionally trusted to enforce the clients security level.
In a conventional MQTT system with a broker that only offers one type of connection, the clients are always aware of the achieved security goals when it comes to the connection of other clients, since each client has the same committed conditions which can then also be expected from other clients. In the case the broker only allows TCP connections the publisher sends the MQTT message in plaintext to the broker, which then distributes the message in plaintext to the subscribers. Both publisher and subscriber have an insecure connection to the broker and therefore do not expect any security measures to be implemented on the path from client to client by any means.
In the case of TLS enforcement of the broker, a publisher sends the MQTT message over a secure communication channel to the broker. The broker in turn forwards the message to the corresponding subscribers also over a secure communication channel. TLS thereby ensures the security goals authentication, confidentiality and integrity of the message during the communication of client and broker. All participating clients therefore can expect these security goals to be continuously guaranteed, i.e. from the moment the publisher sends the message to the broker and then from the broker to the moment the subscriber receives it.

Due to the Client-to-Broker architecture in MQTT, we cannot speak of end-to-end security when we mean that an MQTT message is secured in terms of authentication, confidentiality and integrity continuously on all paths throughout the network, i.e., from publisher to broker and from broker to subscriber. Therefore we introduce a new term, called Client-to-Client security. Client-to-Client security in MQTT, by definition, can only be guaranteed under the assumption of a trusted and non-compromised broker, as the broker always acts as an intermediary between the clients. 
With the introduction of multiple Listeners, the assurance of having Client-to-Client security changes. Assuming a broker offers both TCP and TLS for the connection establishment with clients, then it can no longer be ensured that a message sent or received by a TLS client has actually been secured by measurements for authentication, confidentiality and integrity continuously throughout all paths through the network, since TCP clients can also interact with TLS clients, as previously discussed in Section \ref{sec:hybridproblem}. 
This security problem occurs on the interaction of a TLS publisher with a TCP subscriber (as depicted in case 0 of Fig.~\ref{fig:my_label}) or a TCP publisher with a TLS subscriber (case 3 of Fig.~\ref{fig:my_label}) by the help of the broker, the message is transmitted insecurely over the network at least one time and therefore violates the security goal of Client-to-Client security. 

Thus, we speak of broker-mediated client-to-client security for the remainder of this work.

\subsection{Enforcement of the same level of client-to-client security between clients by an enhanced broker}
For an enforcement by a trustworthy broker, it is upfront important for the broker to detect a conflict between the interests of a publisher and the ones of the current and future subscribers.  
First, the broker needs to identify and retain the security level of the client by inspecting incoming message on publish requests (\texttt{PUBLISH}) or subscription requests (\texttt{SUBSCRIBE}).
Second, the broker needs to detect the conflicts between the security levels and act accordingly on every message forwarding action. 
We enhanced the broker to check the MQTT protocol field \texttt{User Properties}, in which our modified client can signal its wish for enforcement of its own security level beyond the broker. 
While this requires a small modification on the client side, one could also work with unmodified clients and just assume their intention for a certain security level from their current connection type. 
For example messages like \texttt{PUBLISH} or \texttt{SUBSCRIBE} when received over a secured connection, e.g. TLS, would result in the broker assuming that they want to have an enforcement of their security level beyond the broker.
The latter would then only require the clients to switch from a secure to a non-secure connection and vice-versa depending on their desired enforcement.

When a broker receives a \texttt{PUBLISH} message the broker must decide whether the message may be forwarded. 
We consider the following cases based on the MQTT protocol and TLS or plain TCP connected clients:

\begin{enumerate}
\item \textbf{Securely connected publisher sends message and requests to enforce Client-to-Client Security}: \newline
If the MQTT publisher is connected via TLS and has informed the MQTT broker that Client-to-Client security should be enforced by including the respective \texttt{User Properties} in the \texttt{CONNECT} or \texttt{PUBLISH} message, then the broker checks whether the subscribers meet the requirements of Client-to-Client security, i.e. whether they have implemented mechanisms for securing authentication, confidentiality and integrity, before the \texttt{PUBLISH} message is forwarded to the subscribers. 
If a subscriber fulfils these requirements, e.g. when the subscriber is also connected to the broker via TLS, the broker forwards the \texttt{PUBLISH} message to this subscriber.
All subscribers that do not meet the requirements e.g. if they are connected via plain TCP, will be denied access to this \texttt{PUBLISH} message.

\item \textbf{Non-Secure Publisher sends message and subscriber has requested to enforce Client-to-Client Security}: \newline
If the MQTT publisher is connected via plain TCP, the broker checks whether there are any MQTT subscribers that have the requirement to enforce Client-to-Client security and are connected to the broker via TLS. 
The \texttt{PUBLISH} message will thus not be delivered to these clients.
All other subscribers --all TCP subscribers and the TLS subscribers that have not specified the requirement for Client-to-Client security-- will receive the message.
\item \textbf{In all other cases}, the MQTT \texttt{PUBLISH} message can be directly forwarded to the subscriber in the usual way. 
\end{enumerate}

\subsection{Possibility to extend to the security goals}
Note, it is possible to define more fine-grained security goals. 
For the sake of simplicity this work only discusses two distinct level of security, secured (e.g. via TLS) and non-secured (e.g. plain TCP) connections.
However, our approach can be extended to contain any types of security parameters which would allow to be compared against each other in order to let the broker derive  a security level that is sufficient for both clients.
For example, one could be technically far more specific what type of TLS protection is required by specifying the exact list of TLS cipher suites to be used, e.g. specify that ephemeral Diffie-Hellman shall be used with an RSA algorithm using a certain key length and a certain hashing algorithm like \texttt{TLS\_ECDHE\_ECDSA\_WITH\_AES\_256\_GCM\_SHA384}.
This could be used to provide integrity of messages only via choosing non-encrypting TLS ciphersuites. But it could also be extended to take into account the level of security protection by other means, e.g. integrity protection by signatures on the message level directly~\cite{PoehlsESIOT2015}, but it seems to be preferably to favour ECDSA over RSA for less resource consumption~\cite{2016BauerStaudemeyerPoehlsFragkiadakisICICS} on IoT devices. 
Another extension could be to segregate communication by network location, e.g. from outside a trusted network, or when transmitted via VPN.
This is left as future work, but our framework is extensible in this respect. 
Sole requirement is that the broker remains able to compare the security goals to identify the minimum level to make a decision which messages to forward to whom in order to achieve client-to-client security. 

\section{Broker-mediated enforcement in relation to other general security concepts}
\label{sec:discussion}
We have presented an enhancement of the broker, that is extended to be trusted to only selectively forward messages in order to uphold a requested security level by one corresponding client (publisher or subscriber) also beyond the broker to other corresponding clients.
We have termed this broker-mediated client-to-client security in order to highlight the following: i) each corresponding client requests a security level to be enforced beyond the broker to another client and ii) the broker is doing the enforcement and thus needs to be trusted beyond just forwarding messages to all interest correspondents.
In the following we want to briefly highlight that the client could remain unchanged, how our concept relates to other approaches and discuss the overheads of our enhancements. 

\subsection{Compatibility to existing MQTT clients without changing}
\label{sec:nochange}
The good news is that the enhancement could be used without the need to adjust or update MQTT clients, just the broker. 
Of course the clients would then not be able to signal their requested protection level freely, as they are not aware of the additional MQTT flags/headers.
However, one could implement default or even complex policies for devices or topics or use the usage of TLS secured connections as an indicator which level the client wants to uphold, e.g. if the subscriber connects with extra burden via TLS it could be assumed that it wants to listen only to authentic incoming messages coming from secured connections.
For increased interconnection, one could also assume that legacy clients would not want to uphold their own security level beyond the broker, leaving the problem still unsolved for legacy clients.

\subsection{Comparison to other approaches}
\subsubsection{Multilateral security}
The notion "multilateral security" was already established in the 1990s \cite{Rannenberg1993} and early 2000s \cite{Rannenberg2001,Pfitzmann2001,CLAU2001205} and is defined as follows: \textit{"Multilateral security considers different and possibly conflicting security requirements of different parties and strives to balance these requirements. It means taking into consideration the security requirements of all parties involved. It also means considering all involved parties as potential attackers"}~\cite{Rannenberg2001}. 
In a later publication~\cite{Pfitzmann2001}, the term was narrowed down to the following requirements:
\begin{enumerate}
	\item Each party has its particular protection goals. 
	\item Each party can formulate its protection goals.
	\item Security conflicts are recognized and compromises negotiated.
	\item Each party can enforce its protection goals within the agreed compromise.
\end{enumerate}
Besides, it is also clarified that the security properties must be dynamically adaptable \cite{Pfitzmann2001} and that "\textit{Multilateral Security does not necessarily enable every participant to enforce all of her individual security goals, but at least it provides transparency of the security of an action for all parties involved}" \cite{CLAU2001205}.
So, while the definitions and formulations of the various authors differ, they basically all state the same essential requirements. 
Our solution achieves the requirements 1-3, as each client can identify its security goals and the broker recognises conflicts between these stated goals and if possible enforces them. 
When discussing requirement number 4 one could treat the combination of a client and a trusted and enhanced broker as a single party. Broker and publisher can then enforce the security level desired by the publisher beyond the broker towards other subscribers; and broker and subscriber together can enforce it towards publishers.
If one would treat the broker as an individual party in the above meaning, then requirement 4 can not be fulfilled by publishers nor subscribers alone and thus requirement no. 4 could not be met.

\subsubsection{End-to-end security}
The term end-to-end security applied to the publisher subscriber interaction would mean that the broker is only able to break availability by maliciously dropping messages, but the clients would be capable of upholding confidentiality and integrity~\cite{PoehlsESIOT2015,2016BauerStaudemeyerPoehlsFragkiadakisICICS} of messages exchanged even if the broker is malicious and reads or modifies messages. 
This would allow clients to put less trust into the broker, e.g. broker could be honest (forward messages) but be curious (break confidentiality) or even actively tamper with the message as end-to-end means that the client on the other end of the communication will notice. 
This is different to the broker-mediated client-to-client security as it requires a more trusted and enhanced broker.
However, client-to-client removes the burden of having to maintain individual keys and trust relation among clients. 

\subsubsection{Other network security notions}  
As discussed in the related work section, there are approaches to place a transparent TLS proxy and a firewall to secure MQTT brokers. 
Vucnik et al.'s approach that ensures traffic traverses through the proxy and not directly by placing the firewall and proxy in front of a plain MQTT server~\cite{mqtt_firewall:2019}.
As discussed our approach is more versatile and different as we allow the clients to relax their own security level which would not be possible with the transparent TLS proxy.
However, following the idea and borrowing network security terms, one could describe our proposed solution as a firewall inside the broker to separate or segregate the message flows.  
For clarity we decided to not mix existing, clearly defined terminologies and decided to call it broker-mediated client-to-client security.

\subsection{Overheads in prototypical implementation}
In the context of this paper, we developed a first prototype of an MQTT broker that, as described in Section \ref{sec:ouridea}, takes into account the desired security levels of the clients and, if necessary, performs operations to enforce them even beyond the broker. For the implementation we chose the programming language Python 3.8.5, as one of our main goals was to design the code as comprehensible as possible, so that other developers can adopt the concepts for their own purposes, e.g. for the extension of existing MQTT brokers such as HiveMQ or Mosquitto. 

In our conducted experiments to determine the resulting overhead of our approach, we considered two different aspects: (1) the overhead per MQTT message and (2) the computational overhead on the side of the broker, which arises from the enforcement process. Our observations regarding the message size are the following: (1) if the broker implicitly determines and enforces the desired security level of the clients based on their connection types (secured or non-secured), no additional information has be to included in the various MQTT messages and (2) if the MQTT clients should be able to indicate their desired security levels to the broker, it requires a minimum of 7 additional bytes (in the \texttt{User Properties} field of the MQTT message, which was introduced in MQTT version 5). This User Property field either has to be added once when the connection to the broker is established (i.e. in the \texttt{CONNECT} message) to define the security-level for all subsequent \texttt{PUBLISH}/\texttt{SUBSCRIBE} messages or alternatively it can also be added in each \texttt{PUBLISH}/\texttt{SUBSCRIBE} message individually to enable an even more fine-grained definition of security levels. With an extension to the current MQTT specification i.e. with the introduction of a new, dedicated identifier for security-levels in the MQTT message, we would even be able to reduce the overall message overhead to 2 additional bytes (1 byte for the identifier and 1 byte for the security-level).

In order to determine the resulting computational overhead, we measured the time it takes on the broker-side to forward an incoming \texttt{PUBLISH} message to a subscriber. The time is measured from the receipt of the message to the transmission of the message to the subscriber. We measured this for normal broker behavior, i.e. messages are forwarded without any enforcement of security levels beyond the broker, as well as for our extended broker behavior i.e. security levels of the clients are enforced beyond the broker. Our 500 simulation runs show that there are no significant performance differences between these two approaches (median 109ms for the normal behavior and 111ms for the extended behavior).

Here, it should be noted that this is only a prototype, which focuses purely on functionality and does not implement any optimizations. It is therefore conceivable that there may well be greater differences in the case of performance optimizations. In the current state of our research, however, we could not observe any significant performance differences.

\section{Conclusion and future work}
\label{sec:conclusion}

In this paper we have shown the importance of security aspects in publish/subscribe based message queuing systems and pointed out the existence of a fundamental problem in today's systems if they are deployed allowing mixed levels of security in their connection to the clients.
We have exemplified the problem, as well as our solution, with real implementations of TLS and plain TCP connections to an MQTT broker. 
The resulting problem of clients being unable to control the enforcement of security levels beyond the broker can either be solved by strictly isolating non-secured from secured clients, or it requires all devices the burden of constantly being connected over TLS.
The former hinders a multitude of possible communication flows that might be totally acceptable and intended. For the latter, we (and other works) have demonstrated an impact on smaller devices which makes it impossible in many environments to enforce TLS on all clients.

Even though there are many available security mechanisms that address these problems, there are still more than 150,000 MQTT systems freely accessible via the Internet\footnote{https://www.shodan.io, search query 'mqtt', Accessed: 17.01.2021}.


There are reasonable justifications for an MQTT based system to neglect the usage of those security mechanisms, especially the usage of TLS. However, TLS is able to mitigate many and most importantly the most severe attack vectors. We have found that one of the main reasons for neglecting the usage of TLS, is the immense negative impact of TLS on the duration of the connection establishment between the MQTT client and the MQTT broker - the connection establishment using plain TCP takes 10ms on average, while the TLS connection establishment takes 3208 ms. This performance loss occurs, as handling the additional computational overhead of TLS often poses a challenge for the resource constrained IoT devices. 
As a response, we have presented our broker-mediated enforcement to make the use of TLS in MQTT-based systems more fine-grained and thus the extra effort spent for TLS more tolerable. 
In our concept, which we have prototypically implemented, an MQTT broker can be configured to accept multiple connection types, such as TCP and TLS, simultaneously. As a result, TLS can be enforced exclusively on devices that are in need of TLS protection, rather than enforcing it on all devices, which currently is the state-of-the-art. 
Or it can be enforced for certain topics or messages only.
In this paper, we have subsequently presented and defined a new security model for message queuing publish/subscribe systems that follows the approach of multilateral security. 
Our enhanced broker enforces what we termed Client-to-Client security, i.e. a continuous protection of authentication, confidentiality and integrity of the message on all ways throughout the network mediated by the broker from the publisher to the subscriber or vice versa. 
It allows the clients to define their security on a fine-grained level (e.g. per subscribed topic or per published message). 
Furthermore, we have provided a first implementation for MQTT brokers and MQTT clients that successfully realizes multilateral security in MQTT based systems. 
By extending the OASIS MQTT specification, we would be able to improve the method of transferring the client's security goals significantly with respect to the overhead of Multilateral Security (from a minimum of 7 bytes to fixed 2 bytes). Overall, however, with the current state of research of Multilateral Security in MQTT based systems, a new level of security can be achieved. 
\newline
~\newline
%
In future work we will try to integrate the enhanced functionality into established MQTT brokers such as Mosquitto or HiveMQ. Based on these results, further techniques can be developed to improve the overall performance of enforcement of security-levels beyond the broker.

It is also important to clarify whether current smart devices are even capable of supporting the indication of their desired security-levels, i.e. whether these devices support MQTT version 5 and if it is possible to specify the necessary \texttt{User Properties}, or whether the enforcement beyond the broker can only be realized in practice through enforcement based on the connection types; as noted in Sect.~\ref{sec:nochange} our solution would  not require changes on the clients' MQTT or communication stack.

Another aspect to be clarified is how the full trust of the MQTT client into the MQTT broker can be relaxed: Among the possibilities it would be conceivable to use homomorphic encryption to hide specific information of an MQTT message, that should not be revealed to the MQTT broker.

Finally, it is worth investigating whether  well-established light-weight alternatives to the TLS protocol can lead to improvements that are worth the reduced security offered by light-weight cryptography.

\bibliographystyle{alpha}
\bibliography{Bibliography}

\newcommand{\etalchar}[1]{$^{#1}$}
\begin{thebibliography}{CHSvdM16}

\bibitem[ARH17]{security_issues_mqtt}
S.~{Andy}, B.~{Rahardjo}, and B.~{Hanindhito}.
\newblock {Attack scenarios and security analysis of MQTT communication
  protocol in IoT system}.
\newblock In {\em 2017 4th International Conference on Electrical Engineering,
  Computer Science and Informatics (EECSI)}, pages 1--6, Sep. 2017.

\bibitem[BBBG14]{mqtt3_oasis}
A.~{Banks}, E.~{Briggs}, K.~{Borgendale}, and R.~{Gupta}, editors.
\newblock {\em {MQTT Version 3.1.1}}, Woburn, USA, October 2014. Organization
  for the Advancement of Structured Information Standards.
\newblock http://docs.oasis-open.org/mqtt/mqtt/v3.1.1/os/mqtt-v3.1.1-os.html.
  Latest version: http://docs.oasis-open.org/mqtt/mqtt/v3.1.1/mqtt-v3.1.1.html.

\bibitem[BBBG19]{mqtt5:oasis}
A.~{Banks}, E.~{Briggs}, K.~{Borgendale}, and R.~{Gupta}, editors.
\newblock {\em {MQTT Version 5.0}}, Woburn, USA, March 2019. Organization for
  the Advancement of Structured Information Standards.
\newblock https://docs.oasis-open.org/mqtt/mqtt/v5.0/os/mqtt-v5.0-os.html.
  Latest version: https://docs.oasis-open.org/mqtt/mqtt/v5.0/mqtt-v5.0.html.

\bibitem[BSPF16]{2016BauerStaudemeyerPoehlsFragkiadakisICICS}
Johannes Bauer, Ralf~C. Staudemeyer, Henrich~C. Pöhls, and Alexandros
  Fragkiadakis.
\newblock Ecdsa on things: Iot integrity protection in practise.
\newblock In K.-Y.~Lam et~al., editor, {\em Proc. of Information and
  Communications Security (ICICS 2016)}, volume 9977 of {\em LNCS}. Springer,
  Nov. 2016.

\bibitem[CHSvdM16]{tls_scrutinise:2016}
Cas Cremers, Marko Horvat, Sam Scott, and Thyla van~der Merwe.
\newblock Automated analysis and verification of tls 1.3: 0-rtt, resumption and
  delayed authentication.
\newblock In {\em 2016 IEEE Symposium on Security and Privacy (SP)}, pages
  470--485, 2016.

\bibitem[CK01]{CLAU2001205}
S.~Clauß and M.~Köhntopp.
\newblock {Identity management and its support of multilateral security}.
\newblock {\em Computer Networks}, 37(2):205--219, 2001.
\newblock Electronic Business Systems.

\bibitem[DG21]{tls13-proof}
Hannah Davis and Felix G{\"u}nther.
\newblock Tighter proofs for the sigma and tls 1.3 key exchange protocols.
\newblock In Kazue Sako and Nils~Ole Tippenhauer, editors, {\em Applied
  Cryptography and Network Security}, pages 448--479, Cham, 2021. Springer
  International Publishing.

\bibitem[{Esp}20]{espressif_documentation}
{Espressif Documentation}.
\newblock {ESP8266EX Datasheet V6.6}.
\newblock
  \url{https://www.espressif.com/en/support/documents/technical-documents},
  2020.
\newblock Accessed: 15 July 2021.

\bibitem[FOP{\etalchar{+}}16]{bookEngineeringSecureIoTSystems}
Alexandros Fragkiadakis, George Oikonomou, Henrich~C. P{{\"o}}hls, Elias~Z.
  Tragos, and Marcin W\'ojcik.
\newblock Securing communications among severely constrained, wireless embedded
  devices.
\newblock In Benjamin Aziz, Alvaro Arenas, and Bruno Crispo, editors, {\em
  Engineering Secure Internet of Things Systems}. The Institute of Engineering
  and Technology, Oct. 2016.

\bibitem[G{\"o}t14]{mqtt_history:2014}
C.~G{\"o}tz.
\newblock {MQTT: Protokoll für das Internet der Dinge}.
\newblock
  \url{https://m.heise.de/developer/artikel/MQTT-Protokoll-fuer-das-Internet-der-Dinge-2168152.html},
  April 2014.
\newblock Accessed: 03 February 2021.

\bibitem[HBK18]{analysis_of_mqtt_vulnerabilities:2018}
M.~S. {Harsha}, B.~M. {Bhavani}, and K.~R. {Kundhavai}.
\newblock {Analysis of vulnerabilities in MQTT security using Shodan API and
  implementation of its countermeasures via authentication and ACLs}.
\newblock In {\em 2018 International Conference on Advances in Computing,
  Communications and Informatics (ICACCI)}, pages 2244--2250, Sep. 2018.

\bibitem[HMQnd]{hivemq_listeners:2020}
{HiveMQ Documentation - Listeners}.
\newblock
  \url{https://www.hivemq.com/docs/hivemq/4.5/user-guide/listeners.html/}, n.d.
\newblock Accessed: 01 March 2021.

\bibitem[KA16]{iotdevice_classes:2016}
J.~King and A.~Awad.
\newblock {A distributed security mechanism for Resource-Constrained IoT
  Devices}.
\newblock {\em Informatica}, 40:133--143, 01 2016.

\bibitem[{Kas}18]{kaspersky:2018}
{Kaspersky Lab}.
\newblock {Internet of Things: What Is IoT? IoT Security}.
\newblock
  \url{https://www.kaspersky.com/resource-center/definitions/what-is-iot},
  December 2018.
\newblock Accessed: 27 January 2021.

\bibitem[KPS20]{felixkorbihenrich2020}
F.~Klement, H.~Pöhls, and K.~Spielvogel.
\newblock {Towards Privacy-Preserving Local Monitoring and Evaluation of
  Network Traffic from IoT Devices and Corresponding Mobile Phone
  Applications}.
\newblock In {\em {IEEE 3rd Workshop on Internet of Things Security and Privacy
  (WISP 2020) held in conjunction with Global IoT Summit 2020 (GIOTS 2020)}},
  pages 1--6. IEEE, 06 2020.

\bibitem[MK20]{mqtt_survey:2020}
Biswajeeban Mishra and Attila Kertesz.
\newblock The use of mqtt in m2m and iot systems: A survey.
\newblock {\em IEEE Access}, 8:201071--201086, 2020.

\bibitem[P{{\"}}15]{PoehlsESIOT2015}
Henrich~C. P{{\"o}}hls.
\newblock {JSON Sensor Signatures (JSS): End-to-End Integrity Protection from
  Constrained Device to IoT Application}.
\newblock In {\em Proc. of the Workshop on Extending Seamlessly to the Internet
  of Things (esIoT), collocated at the 9th International Conference on
  Innovative Mobile and Internet Services in Ubiquitous Computing (IMIS 2015)},
  pages 306 -- 312. IEEE, Jul. 2015.

\bibitem[Pfi01]{Pfitzmann2001}
A.~Pfitzmann.
\newblock {\em {Multilateral Security: Enabling Technologies and Their
  Evaluation}}, pages 50--62.
\newblock Springer Berlin Heidelberg, Berlin, Heidelberg, 2001.

\bibitem[Ran93]{Rannenberg1993}
K.~Rannenberg.
\newblock {Recent Development in Information Technology Security Evaluation -
  The Need for Evaluation Criteria for Multilateral Security}.
\newblock In {\em Proceedings of the IFIP TC9/WG9.6 Working Conference on
  Security and Control of Information Technology in Society on Board M/S Illich
  and Ashore}, page 113–128, NLD, 1993. North-Holland Publishing Co.

\bibitem[Ran01]{Rannenberg2001}
K.~Rannenberg.
\newblock {Multilateral Security a Concept and Examples for Balanced Security}.
\newblock In {\em Proceedings of the 2000 Workshop on New Security Paradigms},
  NSPW '00, page 151–162, New York, NY, USA, 2001. Association for Computing
  Machinery.

\bibitem[Ske19]{mqtt_popularity:2019}
I.~Skerrett.
\newblock {Why MQTT Has Become the De-Facto IoT Standard}.
\newblock
  \url{https://dzone.com/articles/why-mqtt-has-become-the-de-facto-iot-standard},
  October 2019.
\newblock Accessed: 10 February 2021.

\bibitem[{The}15a]{hivemq_introduction:2015}
{The HiveMQ Team}.
\newblock {Introducing the MQTT Protocol - MQTT Essentials: Part 1}.
\newblock
  \url{https://www.hivemq.com/blog/mqtt-essentials-part-1-introducing-mqtt/},
  January 2015.
\newblock Accessed: 12 April 2021.

\bibitem[{The}15b]{hivemq_mqtt_security_fundamentals:2015}
{The HiveMQ Team}.
\newblock {Introducing the MQTT Security Fundamentals}.
\newblock
  \url{https://www.hivemq.com/blog/introducing-the-mqtt-security-fundamentals/},
  April 2015.
\newblock Accessed: 24 January 2021.

\bibitem[{The}15c]{hivemq_tls_security_fundamentals:2015}
{The HiveMQ Team}.
\newblock {TLS/SSL - MQTT Security Fundamentals}.
\newblock
  \url{https://www.hivemq.com/blog/mqtt-security-fundamentals-tls-ssl/}, May
  2015.
\newblock Accessed: 12 April 2021.

\bibitem[{The}19]{hivemq_connect_establishment:2019}
{The HiveMQ Team}.
\newblock {Client, Broker / Server and Connection Establishment - MQTT
  Essentials: Part 3}.
\newblock
  \url{https://www.hivemq.com/blog/mqtt-essentials-part-3-client-broker-connection-establishment/},
  July 2019.
\newblock Accessed: 14 February 2021.

\bibitem[VvKM19]{mqtt_firewall:2019}
Matevž Vučnik, Aleš Švigelj, Gorazd Kandus, and Mihael Mohorčič.
\newblock Secure hybrid publish-subscribe messaging architecture.
\newblock In {\em 2019 International Conference on Software, Telecommunications
  and Computer Networks (SoftCOM)}, pages 1--5, 2019.

\end{thebibliography}

\end{document}